\documentclass[aps,preprint]{revtex4}
\usepackage{amssymb}
\usepackage{amsmath}

\input{tcilatex}

\begin{document}

\title{Stringy Symmetries and Their High-energy Limit}
\author{Jen-Chi Lee}
\email{jcclee@cc.nctu.edu.tw}
\affiliation{Department of Electrophysics, National Chiao-Tung University, Hsinchu,
Taiwan, R.O.C. \ \ }
\date{\today }

\begin{abstract}
The high-energy limit of stringy symmetries, derived from the decoupling of
two types of zero-norm states in the old covariant first quantized (OCFQ)
spectrum of open bosonic string , are used to reproduce Gross's linear
relations among high-energy scattering amplitudes of different string states
with the same momenta. Moreover, the proportionality constants between
scattering amplitudes of different string states are calculated for the
first few low-lying levels. These proportionality constants are, as
suggested by Gross from the saddle point calculation of high-energy
string-loop amplitudes, independent of the scattering angle $\phi _{CM}$ and
the order $\chi $ of string perturbation theory. The decoupling of
degenerate positive-norm states , which is valid to \textit{all} energy, can
also be derived from these stringy symmetries. The high-energy limit of this
decoupling is found to be consistent with the work of Gross and Manes.
\end{abstract}

\maketitle

In the traditional formulation of a local quantum field theory, a symmetry
principle was postulated, which can be used to determine the interaction of
the theory, e. g., Yang-Mills theories and general relativity. The idea of
\textquotedblleft\ symmetry dictates interaction\textquotedblright\ has thus
become one of the fundamental philosophy to pursue new physics such as GUTs
and supergravities for the last few decades. One of the most important
consequences of these symmetries is the resulting softer ultraviolet
structure of field theories which , in some cases , makes them consistent or
renormalizable quantum field theories when incorporating with quantum
mechanics. In these cases , the Ward identities, the direct consequence of
symmetry on the n-point Green functions of the theory, are intensively used
to remove the unwanted loop divergences in perturbation theory. In contrast
to the local quantum field theory, string theory is very different in this
respect. In string theory, on the contrary, it is the interaction,
prescribed by the very tight quantum consistency conditions due to the
extendedness of string rather than point particle, which determines the form
of the symmetry. In fact, once we settle on the quantum theory of a free
string, the forms of the interactions and thus symmetries of all string
states are fixed by the quantum consistency of the theory. For example, the
massless gauge symmetries of 10D Heterotic string\cite{1} were determined to
be SO(32) or $E_{8}^{2}$ by the string one-loop consistency or modular
invariance of the theory. Some stringy Einstein-Yang-Mills type symmetries
with symmetry parameters containing both Einstein and Yang-Mills index were
proposed in Ref[2]. Being a consistent quantum theory with no free parameter
and an infinite number of states, it is conceivable that there exists huge
symmetry group or Ward identities, which are responsible for the ultraviolet
finiteness of string theory. To uncover the structure of this huge hidden
symmetry group has become one of the most challenging problem ever since the
discovery of string theory.

In 1988 Gross\cite{3} made an important progress on this subject. With the
calculation of high-energy limit of closed string scattering amplitudes for
an arbitrary string-loop order G through the use of a semi-classical, saddle
point technique developed by Gross and Mende\cite{4}, he was able to derive
an infinite number of linear relations among high-energy scattering
amplitudes of different string states with the same momenta. These relations
were shown to be valid order by order and were of the \textit{identical form}
in string perturbation theory. As a result, the high-energy scattering
amplitudes of all string states can be expressed in terms of, say, the
dilaton scattering amplitudes. A similar result was obtained for the open
string by Gross and Manes\cite{5}. However, the physical origin of these
symmetries and thus the meaning of proportionality constants between the
high-energy scattering amplitudes of different string states were unknown to
those authors, and their values were not calculated.

In this letter, we propose an infinite number of stringy Ward identities
derived from the decoupling of two types of zero-norm states\cite{6} in the
OCFQ string spectrum. Thses Ward identities are valid to \textit{all} energy 
$\alpha ^{\prime }$ and to all orders in string perturbation theory since
zero-norn states should be decoupled from the correlation functions at each
order of perturbation theory by unitarity. The simplest example is the
familiar massless \textit{on-shell} Ward identity of string QED. In this
sense, the stringy Ward identities we proposed in this letter serve as a
natural generalization of Ward identity in gauge field theory. As the first
test of these stringy Ward identities, the high-energy limit of them are
used to reproduce Gross's linear relations among high-energy scattering
amplitudes of different string states with the same momenta. Moreover, the
proportionality constants between scattering amplitudes of different string
states are calculated for the first few low-lying levels. We find that these
high-energy proportionality constants are, as suggested by Gross ,
independent of scattering angle $\phi _{CM}$ and the order $\chi $ of string
perturbation theory. However, the proportionality coefficients do depend on
the scattering angle $\phi _{CM}$ through the dependence of momentum at 
\textit{low} energy. To further uncover the group theoretical structure of
these stringy symmetries, it is important to explicitly calculate the
complete set of zero-norm states in the spectrum. Recently, a simplified
method to generate zero-norm states in OCFQ bosonic string was proposed\cite%
{7}. General formulas of some zero-norm tensor states at an arbitrary mass
level were given. Unfortunately, general formulas for the \textit{complete}
set of zero-norm states are still lacking mostly due to the high
dimensionality of spacetime D = 26. However, in the toy 2D string model\cite%
{8}, a general formula of zero-norm states with discrete Polyakov's momenta
at an arbitrary mass level was given in terms of Schur Polynomials\cite{9}.
These zero-norm states were shown to carry the spacetime $\omega _{\infty }$
charges. On the other hand, the complete spacetime symmetry group of toy 2D
string was known to be the same $\omega _{\infty }$, and the corresponding $%
\omega _{\infty }$ Ward identities were powerful enough to determine the
tachyon scattering amplitudes \textit{without }any integration. These
observations in 2D and 26D string theories signal the importance of the
existence of zero-norm states in the OCFQ string spectrum, not shared by
other quantization schemes of string theory, e.g., light-cone quantization.
The advantage of using the decoupling of zero-norm states to derive stringy
Ward identities is that one can avoid the difficult calculation of string-
loop amplitudes. Another one is that the resulting Ward identities are valid
to \textit{all} energy $\alpha ^{\prime }$, in contrast to the high-energy $%
\alpha ^{\prime }\rightarrow \infty $ result of Gross. To further illustrate
this advantage, the decoupling of degenerate positive-norm states\cite{10} ,
which are valid to \textit{all} energy, is also derived from these stringy
Ward identities\cite{7}. This derivation justifies the calculations of two
other independent approaches based on the massive worldsheet sigma-model\cite%
{10} and Witten's string field theory\cite{11}. The high-energy limit of
this decoupling is also found to be consistent with the work of Gross and
Manes\cite{5}.

Let's begin with a brief review of QED Ward identity

\begin{equation}
k_{\mu _{1}}\mathcal{T}^{\mu _{1}\mu _{2}\cdots \mu _{n}}(k_{1}k_{2}\cdots
k_{n})=0,
\end{equation}%
where $\mathcal{T}$ is the\textit{\ off-shell} n-point Green function for n
external photons of polarizations $\mu _{1},\cdots ,\mu _{n}$and momenta $%
k_{1},\cdots ,k_{n}.$Eq. (1) means that the amplitude $\mathcal{T}$\emph{\ }
vanishes if the polarization of one of the external photons is taken to be
logitudinal. Note that eq. (1) holds even off-shell. This seemingly simple
equation, which originated from U(1) gauge symmetry, turns out to be one of
the most far-reaching property of QED. In the old covariant Gupta-Bleuler
quantization of QED, the polarization vector $\epsilon _{\mu }$of photon is
constrained by the covariant gauge condition $%
%TCIMACRO{\FORMULA{k\cdot \varepsilon =0}{k\cdot \varepsilon =0}{evaluate}}%
%BeginExpansion
k\cdot \varepsilon =0%
%EndExpansion
$. One of the three allowed physical polarizations, the logitudinal one $%
%TCIMACRO{\FORMULA{\varepsilon =k}{\varepsilon =k}{evaluate}}%
%BeginExpansion
\varepsilon =k%
%EndExpansion
$, is zero-norm due to the massless condition of on-shell photon. The theory
thus ends up with only two physical transverse propagating modes, and the
logitudinal degree of freedom turns out to serve as the U(1)symmetry
parameter of the theory. In the OCFQ spectrum of open bosonic string theory,
there exists a natural stringy generalization of this zero-norm logitudinal
degree of freedom.They are(we use the notation in Ref[6])

\begin{equation}
\text{Type I}:L_{-1}\left| x\right\rangle ,\text{ where }L_{1}\left|
x\right\rangle =L_{2}\left| x\right\rangle =0,\text{ }L_{0}\left|
x\right\rangle =0;
\end{equation}

\begin{equation}
\text{Type II}:(L_{-2}+\frac{3}{2}L_{-1}^{2})\left\vert \widetilde{x}%
\right\rangle ,\text{ where }L_{1}\left\vert \widetilde{x}\right\rangle
=L_{2}\left\vert \widetilde{x}\right\rangle =0,\text{ }(L_{0}+1)\left\vert 
\widetilde{x}\right\rangle =0.
\end{equation}%
\bigskip While type I states have zero-norm at any space-time dimension,type
II states have zero-norm \emph{only} at D=26. The existence of type II
zero-norm states turns out to be crucial for the discussion in the rest of
this letter. The simplest zero-norm state $k\cdot \alpha _{-1}\mid
0,k\rangle $, $%
%TCIMACRO{\FORMULA{k^{2}=0}{k^{2}=0}{evaluate}}%
%BeginExpansion
k^{2}=0%
%EndExpansion
$ with polarization $k$is the massless solution of eq. (2), which reproduces
the longitudinal photon discussed in eq. (1). A simple prescription to
systematically solve eqs. (2) and (3) for an infinite number of zero-norm
states was given recently in Ref[7]. A more thorough understanding of the
solution of these equations and their relation to space-time $\omega
_{\infty }$ symmetry of toy D=2 string was discussed in Ref[9]. \ 

In the first quantized approach of string theory, the contribution of open
orientable surfaces with Euler number $\chi $ to the four-point open string
scattering amplitude is given by(for our purpose we choose four-point
amplitudes in this letter)

\begin{equation}
\mathcal{T}_{\chi }(k_{i})=g_{c}^{2-\chi }\int \frac{Dg_{\alpha \beta }}{%
\mathcal{N}}DX^{\mu }\exp (-\frac{\alpha ^{\prime }}{2\pi }\int d^{2}\xi 
\sqrt{g}g^{\alpha \beta }\partial _{\alpha }X^{\mu }\partial _{\beta }X_{\mu
})\overset{4}{\underset{i=1}{\Pi }}v_{i}(k_{i})\equiv \langle \overset{4}{%
\underset{i=1}{\Pi }}v_{i}(k_{i})\rangle
\end{equation}%
where $g_{c}$ is the closed string coupling constant, $\mathcal{N}$ is the
volume of the group of diffeomorphisms and Weyl rescalings of the worldsheet
metric, and $v_{i}(k_{i})$ are the on-shell vertex operators with momenta $%
k_{i}$. The integral is over orientable open surfaces of Euler number $\chi $
parametrized by moduli $\overrightarrow{m}$ with punctures at $\xi _{i}$. A
similar formula for the closed string can be easily written down. The string
generalization of eq (1), or the stringy \textit{on-shell} Ward identities
are proposed to be

\begin{equation}
\langle \overset{4}{\underset{i=1}{\Pi }}v_{i}(k_{i})\rangle =0
\end{equation}%
where at least one of the 4 vertex operators corresponds to the zero-norm
state solution of eq. (2) or (3).To illustrate the powerfulness of this
seemingly trivial equation, the D$_{2}$ vector zero-norm state of the second
massive level(spin-three) was calculated to be\cite{12}

\begin{gather}
\{(\frac{1}{2}k_{\mu }k_{\nu }\theta _{\lambda }+2\eta _{\mu \nu \theta
_{\lambda }})\alpha _{-1}^{\mu }\alpha _{-1}^{\nu }\alpha _{-1}^{\lambda
}+9k_{[\mu }\theta _{\nu ]}\alpha _{-2}^{[\mu }\alpha _{-1}^{\nu ]}+12\theta
_{\mu }\alpha _{-3}^{\mu }\}\left\vert 0,k\right\rangle ,  \notag \\
k\cdot \theta =0.
\end{gather}%
Eq. (6) was obtained by antisymmetrizing those terms which involve $\alpha
_{-1}^{\mu }\alpha _{-2}^{\nu }$ in the original type I and type II vector
zero-norm states. If one chooses,say, $v_{1}(k_{1})$\ to be the vertex
operator constructed from zero-norm state in eq. (6), the corresponding
stringy Ward identity of eq. (5) reads\cite{13}

\begin{gather}
(\frac{1}{2}k_{\mu }k_{\nu }\theta _{\lambda }+2\eta _{\mu \nu }\theta
_{\lambda })\mathcal{T}_{2,\chi }^{(\mu \nu \lambda )}+9k_{\mu }\theta _{\nu
}\mathcal{T}_{2,\chi }^{[\mu \nu ]}+12\theta _{\mu }\mathcal{T}_{2,\chi
}^{\mu }=0,  \notag \\
k\cdot \theta =0.
\end{gather}%
where $k_{\mu }\equiv k_{1\mu }$ .We will use 1 and 2 for the incoming
particles and 3 and 4 for the scattered particles. $\mathcal{T}_{2,\chi
}^{\prime }s$ are the second massive level, $\chi $-th order string-loop
amplitudes. For the string-tree level $\chi $=1, the three scattering
amplitudes $\mathcal{T}_{2,\chi }^{\prime }s$ were explicitly calculated and
the Ward identity eq(7) was verified\cite{13}. This inter-particle symmetry
was first discovered\cite{12} by massive worldsheet sigma-model approach and
was recently found to be consistent with the second quantized Witten's
string field theory\cite{11} calculation. At this point, \{$\mathcal{T}%
_{2,\chi }^{(\mu \nu \lambda )},\mathcal{T}_{2,\chi }^{(\mu \nu )},\mathcal{T%
}_{2,\chi }^{\mu }$\}is identified to be the \emph{amplitude triplet} of the
spin-three state. In fact, it can be shown that $\mathcal{T}_{2,\chi }^{(\mu
\nu )}$ and $\mathcal{T}_{2,\chi }^{\mu }$ are fixed by $\mathcal{T}_{2,\chi
}^{(\mu \nu \lambda )}$ due to the stringy Ward identities constructed from
the type I spin-two zero-norm state and another vector zero-norm state
obtained by symmetrizing those terms which involve $\alpha _{-1}^{\mu
}\alpha _{-2}^{\nu }$ in the original type I and type II vector zero-norm
states. $\mathcal{T}_{2,\chi }^{[\mu \nu ]}$ is obviously identified to be
the scattering amplitude of the antisymmetric spin-two state with the same
momenta with $\mathcal{T}_{2,\chi }^{(\mu \nu \lambda )}$.Eq. (7) thus
relates the scattering amplitudes of two different string states with
different spins at the second massive level. Note that eq. (7) is valid
order by order and is \emph{automatically} of the identical form in string
perturbation theory. This is consistent with Gross's argument through the
calculation of high-energy scattering amplitudes. However, it is important
to note that eq. (7) is, in contrast to the high-energy $\alpha ^{\prime
}\rightarrow \infty $ result of Gross, valid to \emph{all} energy $\alpha
^{\prime }$ and does depend on the center of mass scattering angle $\phi
_{CM}$ , which is defined to be the angle between $\overrightarrow{k}_{1}$
and -$\overrightarrow{k}_{3}$, through the dependence of momentum $k$ . To
reproduce Gross's high-energy result and fix the proportionality constants,
which were not dwelt on in Ref[3,5] due to lack of the physical origin of
the proposed high-energy symmetries, one needs to calculate high-energy
limit of eq. (7).

Following Gross\cite{3} and Gross and Manes\cite{5}, high-energy, fixed
angle scattering amplitudes of oriented open strings can be obtained from
those of closed strings calculated by Gross and Mende by using the
reflection principle. First, from eq. (4), one notes that the high-energy
limit $\alpha ^{\prime }\rightarrow \infty $ is equivalent to the
semi-classical limit of first-quantized string theory. In this limit, the
closed string G-loop scattering amplitudes is dominated by a saddle point in
the moduli space $\overrightarrow{m}$. For the oriented open string
amplitudes, the saddle point configuration can be constructed from an
associated configuration of the closed string via reflection principle. It
was also found that the Euler number $\chi $ of the oriented open string
saddle is always $%
%TCIMACRO{\FORMULA{\chi =1-G}{\chi =1-G}{simplify}}%
%BeginExpansion
\chi =1-G%
%EndExpansion
$,where G is the genus of the associated closed string saddle. Thus the
integral in eq. (4) is dominated in the $\alpha ^{\prime }\rightarrow \infty 
$ limit by an associated G-loop closed string saddle point in $X^{\mu }$,$%
\widehat{\overrightarrow{m}_{i}}$ and $\widehat{\xi _{i}}$. The closed
string classical trajectory at G-loop order was found to behave at the
saddle point as\cite{4}

\begin{equation}
X_{c1}^{\mu }(z)=\frac{i}{1+G}\overset{4}{\underset{i=1}{\sum }}k_{i}\ln
\left\vert z-a_{i}\right\vert +O(\frac{1}{\alpha ^{\prime }}),
\end{equation}%
which leads to the G-loop four-tachyon amplitude

\begin{equation}
\mathcal{T}_{G}\approx g_{c}^{1+G}\exp -\alpha ^{\prime }[\frac{1}{4(1+G)}%
\underset{i<j}{\sum }k_{i}k_{j}\ln \left\vert a_{i}-a_{j}\right\vert ]\equiv
g_{c}^{1+G}e^{-\alpha ^{\prime }E_{G}}
\end{equation}%
where E$_{G}$ can be thought of as the electrostatic energy of
two-dimensional Minkowski charges $k_{i}$ placed at $a_{i}$ on a Riemann
surface of genus G. One can use the SL(2,C) invariance of this saddle to fix
3 of the 4 points $a_{i}$, Then the only modulus is the cross ratio $\lambda
=\frac{(a_{1}-a_{3})(a_{2}-a_{4})}{(a_{1}-a_{2})(a_{3}-a_{4})}$, which takes
the value $\lambda =\widehat{\lambda }\approx -\frac{t}{s}\approx \sin ^{2}%
\frac{\phi _{CM}}{2}$ to extremize E$_{G}$ if we neglect the mass of the
tachyons in the high-energy limit. $%
%TCIMACRO{\FORMULA{S=-(k_{1}+k_{2})^{2}}{S=-(k_{1}+k_{2})^{2}}{evaluate}}%
%BeginExpansion
S=-(k_{1}+k_{2})^{2}%
%EndExpansion
$, $%
%TCIMACRO{%
%\FORMULA{t=-(k_{1}+k_{3})^{2}}{t=-\left( k_{1}+k_{3}\right) ^{2}}{evaluate}}%
%BeginExpansion
t=-\left( k_{1}+k_{3}\right) ^{2}%
%EndExpansion
$,and $%
%TCIMACRO{%
%\FORMULA{u=-(k_{1}+k_{4})^{2}}{u=-\left( k_{1}+k_{4}\right) ^{2}}{evaluate}}%
%BeginExpansion
u=-\left( k_{1}+k_{4}\right) ^{2}%
%EndExpansion
$ are the Mandelstam variables. Thus, the high-energy scattering amplitudes
of four open string tachyons at the $\chi $-th order is approximated by\cite%
{4}

\begin{equation}
\mathcal{T}_{\chi }\approx g_{c}^{2-\chi }\exp (-\alpha ^{\prime }\frac{s\ln
s+t\ln t+u\ln u}{2(2-\chi )}),
\end{equation}%
which reproduces the very soft exponential decay e$^{-\alpha ^{\prime }s}$
of the well-known string-tree $\chi $=1 amplitude.

For excited string states, the exponential part of the vertex operator is
the same as that of tachyon. The same saddle point eq. (8) dominates the
scattering amplitude. To estimate the high-energy behavior of these
scattering amplitudes, one needs to know an explicit formula for $\partial
^{n}X$ at the saddle point. By putting charges at $a_{1}$=0, $a_{2}$=1,$%
a_{3} $=$\lambda =\sin ^{2}\frac{\phi _{CM}}{2}$, and $a_{4}$=$\infty $, and
using eq. (8) with z=0, we get\cite{5}

\begin{equation}
\partial ^{n}X\sim \frac{i}{1+G}(-1)^{n-1}(n-1)!(k_{2}+\frac{k_{3}}{\lambda
^{n}}).
\end{equation}%
It is important to see that only polarizations in the plane of scattering
will contribute to the amplitude at high energy. Let's define the normalized
polarization vectors e$_{T}$ and e$_{L}$ with spatial components in the CM
frame contained in the plane of scattering. They are respectively parallel (e%
$_{L}$) and perpendicular (e$_{T}$) to the momentum of the particle. One
has, to leading order in the energy $E$, \cite{5}

\begin{equation}
e_{T}\cdot \partial ^{n}X\sim i(-)^{n}\frac{(n-1)!}{\lambda ^{n}}E\sin \phi
_{CM},n>0;
\end{equation}

\begin{equation}
e_{L}\cdot \partial ^{n}X\sim i(-)^{n}\frac{(n-1)!}{\lambda ^{n}}\frac{%
E^{2}\sin ^{2}\phi _{CM}}{2m}\overset{n-2}{\underset{l=0}{\sum }\lambda ^{l}}%
,n>1;
\end{equation}%
\begin{equation}
e_{L}\cdot \partial ^{n}X\sim 0,n=1,\text{\ \ \ \ \ \ \ \ \ \ \ \ \ \ \ \ \
\ \ \ \ \ \ \ \ \ \ \ \ }
\end{equation}%
where m is the mass of the particle.

We are now ready to take the high-energy limit of eq. (7). First, one notes
that there are (D-1) equations, which relate components of amplitudes of $%
\mathcal{T}_{2,\chi }^{(\mu \nu \lambda )}$ and $\mathcal{T}_{2,\chi }^{[\mu
\nu ]}$ in eq (7) due to the (D-1) independent constants of $\theta _{\mu }$%
. Remember that $\mathcal{T}_{2,\chi }^{\mu }$ is fixed by $\mathcal{T}%
_{2,\chi }^{(\mu \nu \lambda )}$ as has been discussed above. By taking the
high-energy, fixed angle limit of Ref[3,5] and using eqs. (12)-(14), one
ends up with only one nonvanishing component for each scattering amplitude.
They are $\mathcal{T}_{2,\chi }^{TTT}$ for the spin-three state and $%
\mathcal{T}_{2,\chi }^{LT}$ for the antisymmetric spin-two state. Taking $%
\theta _{\mu }$=$\theta _{T}$ and $\eta _{\mu \nu }$=$\eta _{TT}$ in eq(7),
and keeping only the leading order in $E$ , we have

\begin{equation}
2\mathcal{T}_{2,\chi }^{TTT}+\frac{9}{2}\mathcal{T}_{2,\chi }^{LT}=0.
\end{equation}%
Note that $k_{\mu }$ in the second term of eq. (7) has cancelled the mass of
the incident particle in eq. (13), which approaches zero in the high-energy
limit. Eq. (15) gives the proportionality constant between high-energy
scattering amplitudes of two different string states at the second massive
level. This proportionality constant is, as suggested by Gross\cite{3},
independent of the scattering angle $\phi _{CM}$ and the order $\chi $(or G)
of string perturbation theory. \textit{Most importantly, we now understand
that it is originated from D}$_{2}$\textit{zero-norm state in the OCFQ
spectrum of the string!}

To further illustrate this high-energy stringy phenomenon, we go to the
third massive level with four different propagating states. In this level,
it was discovered\cite{7,10,11} that the scattering amplitude of the
spin-two state\cite{14}

\begin{gather}
\lbrack \epsilon _{\mu \nu }\alpha _{-2}^{\mu }\alpha _{-2}^{\nu }+\frac{3}{2%
}k_{\lambda }\epsilon _{\mu \nu }\alpha _{-1}^{\lambda }\alpha _{-1}^{\mu
}\alpha _{-2}^{\nu }-\frac{20}{29}(k_{\mu }k_{\nu }+\frac{1}{5}\eta _{\mu
\nu })\epsilon _{\rho \sigma }\alpha _{-1}^{\mu }\alpha _{-1}^{\nu }\alpha
_{-1}^{\rho }\alpha _{-1}^{\sigma }]\left\vert 0,k\right\rangle ,  \notag \\
\epsilon _{\mu \nu }=\epsilon _{\nu \mu },k^{\mu }\epsilon _{\mu \nu }=\eta
^{\mu \nu }\epsilon _{\mu \nu }=0,
\end{gather}%
can be expressed in terms of those of two higher spin states, the spin-four
and the mixed-symmetric spin-three states, as

\begin{equation}
(\frac{1}{3}k_{\lambda }\epsilon _{\mu \nu }+\frac{1}{2}k_{(\lambda
}\epsilon _{\mu \nu )})\mathcal{T}_{3,\chi }^{\lambda \mu \nu }+(\frac{13}{%
174}k_{\alpha }k_{\beta }\epsilon _{\mu \nu }+\frac{3}{58}\eta _{\alpha
\beta }\epsilon _{\mu \nu })\mathcal{T}_{3,\chi }^{\mu \nu \alpha \beta },
\end{equation}%
where $\mathcal{T}_{3,\chi }^{\lambda \mu \nu }$and $\mathcal{T}_{3,\chi
}^{\mu \nu \alpha \beta }$ correspond to the amplitudes of $\alpha
_{-1}^{\lambda }\alpha _{-1}^{\mu }\alpha _{-2}^{\nu }$ and $\alpha
_{-1}^{\mu }\alpha _{-1}^{\nu }\alpha _{-1}^{\alpha }\alpha _{-1}^{\beta }$
respectively. Note that $\mathcal{T}_{3,\chi }^{(\lambda \mu \nu )}$is fixed
by $\mathcal{T}_{3,\chi }^{\mu \nu \alpha \beta }$ due to the existence of
the spin-three zero-norm state at this mass level. Eq (17) is obtained by
making use of the type I spin-two Ward identity

\begin{gather}
2\theta _{\mu \nu }^{1}\mathcal{T}_{3,\chi }^{\mu \nu }+4\theta _{\mu \nu
}^{1}\widetilde{\mathcal{T}}_{3,\chi }^{\mu \nu }+2(k_{\lambda }\theta _{\mu
\nu }^{1}+k_{(\lambda }\theta _{\mu \nu )}^{1})\mathcal{T}_{3,\chi
}^{\lambda \mu \nu }+\frac{2}{3}k_{\lambda }k_{\beta }\theta _{\mu \nu }^{1}%
\mathcal{T}_{3,\chi }^{\mu \nu \lambda \beta }=0,  \notag \\
\theta _{\mu \nu }^{1}=\theta _{\nu \mu }^{1},k^{\mu }\theta _{\mu \nu
}^{1}=\eta ^{\mu \nu }\theta _{\mu \nu }^{1}=0
\end{gather}%
and the type II spin-two Ward identity

\begin{gather}
3\theta _{\mu \nu }^{2}\mathcal{T}_{3,\chi }^{\mu \nu }+8\theta _{\mu \nu
}^{2}\widetilde{\mathcal{T}}_{3,\chi }^{\mu \nu }+(k_{\lambda }\theta _{\mu
\nu }^{2}+\frac{15}{2}k_{(\lambda }\theta _{\mu \nu )}^{2})\mathcal{T}%
_{3,\chi }^{\lambda \mu \nu }+(\frac{1}{2}\eta _{\lambda \beta }\theta _{\mu
\nu }^{2}+\frac{3}{2}k_{\lambda }k_{\beta }\theta _{\mu \nu }^{2})\mathcal{T}%
_{3,\chi }^{\mu \nu \lambda \beta }=0,  \notag \\
\theta _{\mu \nu }^{2}=\theta _{\nu \mu }^{2},k^{\mu }\theta _{\mu \nu
}^{2}=\eta ^{\mu \nu }\theta _{\mu \nu }^{2}=0.
\end{gather}%
$\mathcal{T}_{3,\chi }^{\mu \nu }$ and $\widetilde{\mathcal{T}}_{3,\chi
}^{\mu \nu }$ in eqs. (18) and (19) correspond to the amplitudes of $\alpha
_{-2}^{\mu }\alpha _{-2}^{\nu }$ and $\alpha _{-1}^{\mu }\alpha _{-3}^{\nu }$
respectively. Eqs. (18) and (19) are generated by two degenerate spin-two
zero-norm states at the third massive level. Similarly, the scattering
amplitude of the scalar state\cite{14}

\begin{eqnarray}
&&(\eta _{\mu \nu }+\frac{13}{3}k_{\mu }k_{\nu })\mathcal{T}_{3,\chi }^{\mu
\nu }+(\frac{20}{9}k_{\mu }k_{\nu }k_{\rho }+\frac{2}{3}k_{\mu }\eta _{\nu
\rho }+\frac{13}{3}k_{\rho }\eta _{\nu \rho })\mathcal{T}_{3,\chi }^{\mu \nu
\rho }  \notag \\
&&+(\frac{23}{81}k_{\mu }k_{\nu }k_{\rho }k_{\sigma }+\frac{32}{27}k_{\mu
}k_{\nu }\eta _{\rho \sigma }+\frac{19}{18}\eta _{\mu \nu }\eta _{\rho
\sigma })\mathcal{T}_{3,\chi }^{\mu \nu \rho \sigma }
\end{eqnarray}%
is also fixed by mixed-symmetric $\mathcal{T}_{3,\chi }^{\lambda \mu \nu }$
and $\mathcal{T}_{3,\chi }^{\mu \nu \rho \sigma }$\cite{7}. This stringy
decoupling phenomenon is, in contrast to the result of Gross, valid to 
\textit{all} energy, and was first discovered \cite{10} by using the massive
worldsheet sigma-model approach. It has also been justified recently by
Witten string field theory\cite{11} and the S-matrix calculation\cite{7}. We
can now take the high-energy limit of eqs. (18) and (19) by making use of
eqs. (12)-(14) to get

\begin{equation}
\mathcal{T}_{3,\chi }^{LL}=\frac{1}{9}\mathcal{T}_{3,\chi }^{TTTT},\mathcal{T%
}_{3,\chi }^{TTL}=-\frac{1}{3}\mathcal{T}_{3,\chi }^{TTTT}.
\end{equation}%
Note that the gauge conditions $k^{\mu }\theta _{\mu \nu }=0$ of eqs. (18)
and (19) drop out in the high-energy limit since $k^{\mu }\theta _{\mu
\lambda }=k^{\mu }\theta _{LL}\sim m\approx 0$. The high-energy limit of
eqs. (17) and (20) then give the scattering amplitudes of the spin-two and
the scalar states to be -$\frac{1}{261}\mathcal{T}_{3,\chi }^{TTTT}$ and $%
\frac{5}{54}\mathcal{T}_{3,\chi }^{TTTT}$ respectively.

We conclude that the physical orgin of the proportionality constants in eqs
(15) and (21) is from the zero-norm states in the OCFQ spectrum. However,
the physical effect of them is still not clear at this point. Perhaps the
stringy symmetries proposed in this letter are being realized in the wrong
vacuum as was strongly suggested by the non-Borel summability of string
perturbation theory\cite{15}. In this case, it is then difficult to give a
physical interpretation of these proportionality constants although knowing
their orgin from zero-norm states is still crucial to uncover the
fundamental structure of string theory. The most challenging problem
remained is the calculation of algebraic structure of these stringy
symmetries derived from solutions of eqs. (2) and (3). Presumably, it is a
complicated 26D generalization of $\omega _{\infty }$ of the simpler toy 2D
string model.

\begin{center}
Acknowledgments
\end{center}

I would like to thank the hospitalities of physics departments of National
Taiwan university and Simon-Fraser university, where most of this work was
done. This work is supported by grants of National Science Council of
Taiwan.\

\end{document}